\documentclass{amsart}
\usepackage{amsmath,amssymb,amsthm}
\usepackage{paralist}
\usepackage{algpseudocode}

\usepackage[colorlinks=true]{hyperref}
\hypersetup{urlcolor=blue, citecolor=red}

\newtheorem{thm}{Theorem}[section]
\newtheorem{pro}[thm]{Proposition}
\newtheorem{lem}[thm]{Lemma}

\theoremstyle{definition}
\newtheorem{defin}{Definition}
\newtheorem{example}{Example}
\newtheorem{algorithm}{Algorithm}

\title[]{Complete Gr\"obner basis for lattice codes}
\author[]{I. \'Alvarez-Barrientos, M. Borges-Quintana}
\address{I. \'Alvarez-Barrientos, M. Borges-Quintana, J. A. Ornella Rodr\'iguez \newline
	Departamento de Matem\'atica, Universidad de Oriente,{Santiago de Cuba, Cuba}}
\email{ismara.alvarez@uo.edu.cu, mijail@uo.edu.cu, joseornella93@gmail.com}
\author[]{ M. A. Borges Trenard}
\address{M. A. Borges Trenard\newline
	 Doctorate in Mathematics Education, Universidad Antonio Nari\~no, Bogot\'a, Colombia}
\email{borgestrenard2014@gmail.com}
\author[]{E. Mart\'inez Moro} 
\address{E. Mart\'inez Moro\newline
	 Institute of Mathematics, University of Valladolid,{Valladolid, Spain}}
\email{edgar.martinez@uva.es}
\author[]{J. A. Ornella Rodr\'iguez}
 
\begin{document}
\maketitle
 \begin{abstract}
	In this work, two algorithms are developed related to lattice codes. In the first one,   an extended complete Gr\"obner basis is computed  for the label code of a lattice. This basis supports all term orderings associated with a total degree order  offering   information about de label code of the lattice. The second one   is a decoding algorithm that uses an extended complete Gr\"obner basis of the label code of the lattice for monomial reduction, this provides  all the lattice vectors that constitute candidates for the solution of the Close Vector Problem for a given vector.
\end{abstract}

\section{Introduction}
Lattices are used in communications for coding over band-limited Gaussian channels and the major complexity associated with a lattice code is
the process of decoding, that is, finding the point of the code
that has the smallest   distance to an arbitrary input
(maximum-likelihood decoding).  That is one of the reasons why decoding within  a lattice is an interesting problem nowadays, closely related to the well-known Close Vector Problem (CVP), see for example~\cite{lat2,lat3} and the references therein. Recently, some Gr\"obner basis techniques have been applied to this type of problems, see \cite{lat1,papermio}.  In this work, an algorithm  is proposed that computes an extended complete Gr\"obner basis of the label code for a given lattice. This basis supports all term orderings associated with a total degree compatible  ordering  and  it provides  great  information about the label code associated to the lattice. With that  extended complete Gr\"obner basis   it is possible, via the   reduction of the monomials that  it provides, to create an algorithm that  obtains all the lattice vectors that are candidates for the solution of the CVP for a given fixed vector. 
 
 The structure of this paper will be as follows. In Section~\ref{sec:Pre}, some preliminaries are shown related to lattices, group codes and their associated ideals. Section~\ref{salgMo_CL} 
 shows and identification that allows us to compute a Gr\"obner basis of the ideal of a given lattice group code that allows decoding with respect the $G$-norm. In Section~\ref{sec:Complete}, we discuss the concept of complete Gr\"obner basis and its relation with the coset leaders of the group code and propose  the concept of extended complete Gr\"obner basis and an algorithm that computes it. Finally, in Section~\ref{sec:final} we provide a decoding algorithm based on the extended complete Gr\"obner basis  as well as some examples.

\section{Preliminaries}\label{sec:Pre}

A \textit{ lattice}  $\Lambda$ is a discrete additive subgroup of  $\mathbb{R}^{n}$. A lattice is spanned by integer linear combinations $\Lambda = \left\{ k_{1}\mathbf v_{1} + \ldots + k_{r}\mathbf v_{r}\,|\,k_{i} \in \mathbb{Z} \right\}$ of a given basis $\{\mathbf v_{1},\ldots,\mathbf v_{r}\}$ ($r<n$), where $r$ is the rank of $\Lambda$. From now on,  we will call  $B$ the matrix whose rows are the elements of a given basis of the lattice, i.e. $\Lambda = \left\{ v = xB \; | \; x\in \mathbb{Z}^{n} \right\}$. The\textit{ determinant of the lattice} $\Lambda$ is given by  $\mathrm{det}(\Lambda)=(BB^{t})^{1/2}$ where $B^t$ denotes the transpose of the matrix $B$. Note that  if  $m  =  n$, then  $\mathrm{det}(\Lambda)=\left| \mathrm{det}(B)\right|$. The \textit{dual lattice} $\Lambda^\star$ is the set of linear functionals on $\Lambda$  which take integer values on each point of $\Lambda$, i.e. $\Lambda^\star =\{\mathbf v\in {\text{span}}(\Lambda)\mid \forall \mathbf x\in \Lambda,\,\mathbf x\cdot \mathbf v\in \mathbb {Z} \}$.

Consider $\mathcal V=\{V_i\}_{i=0}^n$ a nested sequence  of vector subspaces of $\mathbb R^n$, $\{\mathbf 0\} = V_0 \subset V_1 \subset \ldots \subset V_n = {\mathbb R}^n$, such that  $\mathrm{dim}(V_i) = i$ and  $V_i = V_{i-1}\oplus W_i$, for  $1 \leq  i \leq n$. If $\Lambda \subset  V_n$  is an $n$-dimensional lattice, we will call the $i$-th \textit{cross sections} of $\Lambda$ by the nested sequence $\mathcal V$ to   $\Lambda_{V_{i}}=\Lambda \cap V_i$ and $\Lambda_{W_{i}}=\Lambda \cap W_i$, and they also have a lattice structure. We will denote the projection of $\Lambda$ on $W_{i}$  as $P_{W_{i}}(\Lambda)$. 

Given a lattice $\Lambda$ and a nested sequence $\mathcal V$, we  consider the group  $G_{i}(\Lambda)\simeq P_{W_{i}}(\Lambda)/\Lambda_{W_{i}}$ where the isomorphism is considered as abelian groups and  $1\leq i \leq n$.  A  \textit{lattice code} $L$ in $\Lambda$ is just a subgroup of $\prod_{i=1}^n G_{i}(\Lambda)$. In \cite{lat1},  the following  result is proposed to check whether a vector in the lattice is in a code or not.

\begin{pro}[\cite{lat1}]\label{prop1}
Let $\mathbf c\in G =\mathbb{Z}/{g_{1}} \mathbb{Z}  \times \dots  \times \mathbb{Z}/{g_{n}}\mathbb{Z}$ an element in the group $G$  associated to the lattice $\Lambda$ and $L$ a lattice code in $G$. Then $\mathbf c\in L$ if and only if 
$\mathbf c\cdot P(\Lambda)B^{\star^{t}}\in \mathbb{Z}^{s},$ 
where  the rows of $B^{\star}$ generate $\Lambda^{\star}$ the dual lattice of $\Lambda$, $s$ is the number of generators of $V^{\star}$, and $P(\Lambda) =  \mathrm{diag}(\mathrm{det}(P_{W_{1}}(\Lambda)),\ldots,  \mathrm{det}(P_{W_{n}}(\Lambda)))$.
\end{pro}
Note that the above condition in the proposition can be seen as a check thus, we will call \textit{parity check matrix} of the code  to the matrix $H=P(\Lambda)B^{\star^{t}}$.
Let $\Lambda$ be a lattice, and $\{\mathbf v_1,\ldots,\mathbf v_r\}$
be a generator set for $\Lambda$, $L$ a label code, and
$C(\Lambda) = \mathrm{diag} (\det (\Lambda_{W_{1}}),\ldots,$
$\det (\Lambda_{W_{n}}))$. In a given coordinate system,
$\mathbf v \in \Lambda$ can be written as 
\begin{equation}\label{eq:morphism}
\mathbf v= \mathbf k\cdot C(\Lambda)+\mathbf c\cdot P(\Lambda)
\end{equation}
where  $\mathbf c\in L$ and $\mathbf k \in \mathbb{Z}^{n}$; see \cite{lat2} for a proof. We
define  $\Phi:\,\Lambda\,\mapsto L$ as the morphism assigning to any 
$\mathbf v\in \Lambda$ the codeword $\mathbf c\in L$ given in Equation~(\ref{eq:morphism}). The set
$\{\Phi(\mathbf v_1),\ldots,\Phi(\mathbf v_r)\}$ is a generator set for the lattice code $L$,
see \cite[Section 3.D, p.~828]{lat2} for a proof.

\subsection{Ideals associated with  group codes}
From now on, we will assume that we have a group  
$G =\mathbb{Z}/{g_{1}} \mathbb{Z}  \times \dots  \times \mathbb{Z}/{g_{n}}\mathbb{Z}$ and
$L$ a lattice  code over $G$, we will denote it as $L\trianglelefteq G$. We will develop our computations over the ring of polynomials $\mathbb K[x_1,x_2,\ldots, x_n]$ where $\mathbb K$ is a field. Note that any field can be used since all the relevant information will be located in the exponents of  monomials, thus usually $\mathbb K=\mathbb F_2$ will be chosen. In this section we will introduce a setting  that provides a general insight of the approach in
\cite{art3,art4} for binary codes, the one in \cite{edgire} for linear codes over (not necessary binary) finite fields  and  for codes over ${\mathbb Z}_m={\mathbb Z}/m{\mathbb Z}$, and in
\cite{lat1} for label codes of lattices.

Depending on the context, we will consider  an element as a integer
or as an element in the group $G$. For instance, in the monomial $x^\mathbf{a}=x_1^{a_1}x_2^{a_2}\cdots x_n^{a_n}$, $\mathbf{a}$ is a vector
in $\mathbb Z_{\geq 0}^n$, while for $\mathbf{a} \in G$, each component $a_i$
is the corresponding element in $G_i$. Abusing the notation, for $\mathbf{a}\in G$, $x^\mathbf{a}$ is the
monomial such that the exponent of each variable is the corresponding
$a_i$ as an integer number, $0\leq a_i\leq g_i-1$. This abuse of notation
can be solved introducing two cross characteristic functions as in
\cite{edgire}, but we will avoid it to get a clearer notation. The support of $\mathbf{a}$ is defined as
$\mathrm{supp}(\mathbf{a})=\{i\in 1,\ldots, n\mid a_i\neq 0\}$, and if $i\in\mathrm{supp}(\mathbf{a})$ we will   also say that $x_i \in \mathrm{supp}(x^\mathbf{a})$.
Given a code $L\trianglelefteq G$, there are  three ways of associating an ideal in the polynomial ring $\mathbb F_2[\mathbf x]=\mathbb F_2[x_1,x_2,\ldots, x_n]$  to  $L$ given in the following three equations.
 
\begin{equation}
I_\equiv({L})=\langle\{x^\mathbf{a} - x^\mathbf{b}\mid  \mathbf{a},\mathbf{b}\in
     {\mathbb Z}_{\geq 0}^n,\, \mathbf{a}-\mathbf{b} \in L\}\rangle \subseteq \mathbb F_2[\mathbf x],
\end{equation}
\begin{equation}
I_1({L})=\langle \{x^\mathbf{a} - 1\mid \mathbf{a} \in L\}\cup
      \{x_i^{g_i}-1\mid i=1,\ldots,n\}\rangle \subseteq  \mathbb F_2[\mathbf x],
\end{equation}
and if we know a set of generators of  $L=\langle \{\mathbf{a}_1,\ldots,\mathbf{a}_k\} \rangle$, 
\begin{equation}\label{eqIL2}
I_2(L)=\langle \{x^{\mathbf{a}_i} - 1\mid  i=1,\ldots,n\}\cup
      \{x_i^{g_i}-1\mid i=1,\ldots,n\}\rangle \subseteq \mathbb F_2[\mathbf x].
\end{equation}
The first one  is a natural way of introducing a binomial ideal
associated with $L$ using the equivalence relation that determines
$L\trianglelefteq G$. It was used in \cite{art4} for the ideal associated with a linear
code\footnote{In order to  consider the non-binary case in this definition,
the linear code ${\mathcal C}$ has to be interpreted over
${\mathbb F}_q^n$, with $q=p^m$ and $p$ a prime number, as the
isomorphic monoid structure of the corresponding code over
${\mathbb F}_p^{mn}$.} and  this was also introduced in  in \cite{edgire}
for codes over ${\mathbb Z}_m$. On the other hand, $I_2(L)$
was defined in \cite{art3,edgire} where it was shown that
it is the same ideal as $I_\equiv({L})$. In \cite{lat1},
$I_1(L)$ was used to define the ideal associated with a lattice
by means of the label code of the lattice.
It was proved in  \cite{papermio} that for a given lattice code $L\trianglelefteq G$, the three ideals above coincide, i.e.  $I_1({L})=I_2({L})=I_\equiv({L})$.
Thus, from now on,  we will that ideal  by $I({L})$.

Let $\left[ X \right]$ be  the set of monomials in the  variables $x_1,\ldots,x_n$ and 
$\prec$ an admissible term order on $\left[ X \right]$.  Given a polynomial $f\in \mathbb F_ 2 [\mathbf x]$,  $T_{\prec}(f)$ will denote
the maximal term of the polynomial $g$. The set of maximal terms of
a set of polynomials $F$ with respect to $\prec$ is
$T_{\prec}\{ F \}=\{ T_{\prec}(f) \mid f \in F \setminus \{ 0 \} \}$.
The set of maximal terms of an ideal $I$ is denoted by $T_{\prec}(I)$
and is the set of maximal terms of the polynomials that belong
to the ideal. Let $N_{\prec}(I) = \left[ X \right] \setminus T_{\prec}(I)$,  the $\mathbb F_2$-vector space $\mathrm{span}_{\mathbb F_2}\left( N_{\prec} \left( I\right) \right)$ with basis $N_{\prec}(I)$ fulfills the following (see for example  \cite{ahuAL94} for a proof),
\begin{enumerate}
	\item $\mathbb F_2[\mathbf x] = I \oplus \mathrm{span}_{\mathbb F_2}(N(I))$.
	\item For all $f \in \mathbb F_2[\mathbf x]$ there exists a unique polynomial
	in $\mathrm{span}_{\mathbb F_2}(N(I))$, denoted by $ \mathrm{Can}(f, I)$ \index{canonical form
		of $f$|textit}, such that $f -\mathrm{Can}(f,I) \in I$; moreover
	\begin{enumerate}
		\item $\mathrm{Can}(f, I) = \mathrm{Can}(g, I)$ if and only if $f - g \in I$.
		\item $\mathrm{Can}(f, I) = 0$ if and only if $f \in I$.
	\end{enumerate}
	\item For each $f \in \mathbb F_2[\mathbf x]$, $T(\mathrm{Can}(f, I)) \preceq T(f)$.
	\item There exists an isomorphism of $\mathbb F_2$-vectorial spaces
	between $\mathrm{span}_{\mathbb F_2}(N(I))$ and $\mathbb F_2[\mathbf x]/I$ (this isomorphism associates
	$\mathrm{Can}(f, I)$ with the class of $f$ modulo $I$).
\end{enumerate}
Let $\mathcal G\subset I$ be a set of polynomials generating the ideal $I$.
Then, $\mathcal G$ is a \emph{Gr\"{o}bner basis} w.r.t. $\prec$ of the ideal $I$  if $T_{\prec}\left\{\mathcal G\right\}$ generates $T_{\prec}(I)$. The  Gr\"{o}bner basis $\mathcal G$ of $I$ is called a
\emph{reduced Gr\"{o}bner basis}  w.r.t. $\prec$  if for all $f \in \mathcal G$,
(i) $T_\prec(f)$ is not a multiple of any $ \in T_\prec(\mathcal G) \setminus \{T_\prec(f)\}$,
	(ii) the leading coefficient of $f$ is 1,
and	(iii) $f = T_\prec(f)-\mathrm{Can}(T_\prec(f),I)$.

\section{M\"{o}ller's algorithm for lattices and decoding}\label{salgMo_CL}
 
We can associate a M\"oller's like  algorithm to a group lattice $G$ using its  additive
monoid structure. For a detailed description of M\"oller's algorithm  on this setting see \cite{art4}, the interested reader can  check in \cite{moraSPES2,mora} the
theoretical foundation of M\"{o}ller's algorithm and a
description of several different contexts where it has been
successfully used. 
For setting the algorithm 
the main objects  will be the following. Let $[X]$ denote the set of monomials in the variables $x_1,x_2, \ldots, x_n$, 
the injective linear morphism
$$\xi:\, \left[X\right]\mapsto \mathbb{R}^{n}$$ will map each
monomial to a vector in $\mathbb{R}^{n}$ as follows:
\begin{enumerate}
\item let $\mathbf e_{i}$ be the $i$-th coordinate vector of $G$, that is, $\mathbf e_i=(e_{ij})_{j=1}^n$ where 
$e_{ii}=1_{{\mathbb Z}_{g_{i}}}$ and $e_{ij}=0$ if $i\neq j$;
\item $\xi(x_{i}) = \mbox{frac}(\mathbf e_{i}P(\Lambda)B^{\star t})$, that is,
$\xi(x_{i}) = \mbox{frac}(\mathbf e_{i}H)$, where $\mbox{frac}(\cdot)$
represents the fractional part between 0 and 1 of the real number;
\item $\xi(\prod^{n}_{i=1}x_{i}^{m_{i}})
= \mbox{frac}(\sum^{n}_{i=1}(m_{i}\;\mbox{mod}\;g_{i})\xi(x_{i}))$,
that is, 
$$\xi(\prod^{n}_{i=1}x_{i}^{m_{i}})
= \mbox{frac}((\sum^{n}_{i=1}m_{i}\mathbf e_{i})H)$$
\end{enumerate}	
We will denote by $\psi:\,\left[X\right]\mapsto G$ to the morphism of
monoids such that $\psi(x_i)=\mathbf e_i$ and it is extended in a 
natural way to all the monomials as $\psi(u)=\psi(\prod^{n}_{i=1}x_{i}^{m_{i}})
=\sum^{n}_{i=1}m_{i}\mathbf e_{i}$. It is clear that   $\xi(u)= \psi (u)H$.

Given a lattice  code $L\trianglelefteq G$, the equivalence relation $R_{L}$  in the quotient $G/L$  can be translated to the monomials in $\left[X\right]$ as follows: 
$$x^\mathbf{a},x^\mathbf{b}\in \left[X\right], \; x^\mathbf{a}\equiv_{L}x^\mathbf{b}\hbox{  if and only if }
(\psi(x^\mathbf{a}),\psi(x^\mathbf{b}))\in R_{L},$$ that is,  by Proposition~\ref{prop1}
 if and only if $(\psi(x^\mathbf{a})-\psi(x^\mathbf{b}))H \in \mathbb{Z}^{n}$.
In other words, two monomials $x^\mathbf{a},x^\mathbf{b}\in \left[X\right]$ are related   if the fractional parts of $\psi(x^\mathbf{a})$ and $\psi(x^\mathbf{a})$
are equal.

 It is clear that $I(L)$
 is an ideal that determines a finite dimensional quotient algebra (zero-dimensional
 ideal), and that the $\mathbb F_2$-linear space  $\mathrm{span}_{\mathbb F_2}(N(I(L)))$ can be  represented, using the arguments above,  by a $\mathbb F_2$-linear space with an effective function dealing with the linear dependency, since the mapping $\xi$ provides a unique canonical form for each element in the space. Thus, once
 an ordering $\prec$ among the terms in $[X]$ has been fixed, an instance of  M\"oller's algorithm will provide us $\mathcal G_\prec(L)$ the  reduced Gr\"obner basis of the ideal $I(L)$ w.r.t. the ordering  $\prec$.

\subsection{Decoding}\label{ssec:deccl}

Consider now $\prec_{tdc}$  a total degree compatible ordering on $[X]$. Given $\mathbf{a}=(a_{1},\ldots,a_{n})$ an element in in a group code  $G$, we define the $G$-norm of $\mathbf{a}$ (see \cite{lat1})  as 
\begin{equation}
\left\|\mathbf{a}\right\|_{G}=a_{1}+\ldots+ a_{n}\in\mathbb R. 
\end{equation}

\begin{lem}
Let $\mathbf{a},\mathbf{c} \in G$ and $\mathbf{c}\prec_{tdc} \mathbf{a}$. If $\mathbf{a}$ is the nearest codeword to $\mathbf{c}$ w.r.t. $\prec_{tdc}$, then  $\mathbf{a}$ is a closest codeword to $\mathbf{c}$ w.r.t. the $G$-norm. 
\end{lem}

{ \begin{proof} This is Lemma 16 in \cite{lat1}. Note that in their proof, the authors consider always the closest as the smallest w.r.t. $\prec_{tdc}$, that is the reason why they claim it is the closest codeword to $\mathbf{c}$ w.r.t. the $G$-norm, but it could be the case of several codewords attaining the smallest $G$-norm (see Example~\ref{ex:1} in this paper), what is true following the proof in  \cite{lat1} is that $\mathbf{a}$ is one of those closest codewords. 
	\end{proof}
}

Let  $\varphi$ be a morphims relating a monomial and a group element in  $G$, 
$\varphi:  [X]  \rightarrow G$ where 
$x^{\mathbf{a}} \mapsto (a_{1}\mathrm{mod}\, g_{1},\ldots,a_{n} \mathrm{mod}\, g_{n})$.

\begin{thm}\label{teop}
Let  $L\trianglelefteq G$,  and $\mathcal G_{\prec}({L})$ a Gr\"obner basis of $I(L)$ w.r.t. $\prec_{tdc}$. Let  $\mathbf{a}\in G$ be an arbitrary vector and  $x^{\mathbf{a}}$ its associated monomial,  and  $\mathbf{e}=\varphi(\mathrm{Can}(x^{\mathbf{a}},\mathcal G_{\prec}({L})))$. Then $\mathbf{c}=\mathbf{a}-\mathbf{e}$ is  one of the  closest codewords in  $L$ to $\mathbf{a}$  w.r.t. $\left\| . \right\|_{G}$.
\end{thm}
 {\begin{proof}
 		The result follows directly from the lemma above and the reduction process of a Gr\"obner basis.
 	\end{proof}
 
}
 
This result generalize those in  \cite{art3,art4} in the context of group codes. On the other hand,  those vectors $\mathbf{a}\in G$ with smallest weight w.r.t.  $\left\| . \right\|_{G}$ in a coset of  $G/L$ (coset leaders) are in correspondence with those monomials of smaller degree   $x^{\mathbf{a}}$  w.r.t. $\equiv_{L}$.

\section{Complete Gr\"obner basis}\label{sec:Complete}

A \textit{complete Gr\"obner basis} (CGb) w.r.t. the ordering $\prec$ associated to the group code $L \trianglelefteq G$, is a Gr\"obner basis for $L$ such that if we permute  the indeterminates keeping the underlying ordering it is still a
 a Gr\"obner basis for $L$ w.r.t. $\prec$. The  \textit{complete reduced Gr\"obner basis} (CrGb) w.r.t. the ordering $\prec$ is the union of all the reduced Gr\"obner basis associated to $L$ taking into account the permutations of the variables, note that  the CrGb is indeed a CGb also and that it is unique. It is clear that a binomial is in the CrGb if and only if  it belongs to a reduced Gr\"obner basis for some ordering on the indeterminates. This concept is somehow related to the  degree compatible Gr\"obner fan in \cite{fan}.

The CrGb   is given by all the binomials of the form  $x^\mathbf{w}-x^{\mathbf{w}^\prime}$, $\mathbf{w},{\mathbf{w}^\prime}\in \mathbb Z_{\geq 0}^n$,  such that  $x^{\mathbf{w}^\prime}$ is the canonical form of  $x^\mathbf{w}$. Constructing such a set has a great computational cost and it is difficult to accomplish  since not all the coset leaders in $G/L$ for the $G$-norm can be seen as a canonical form for an ordering in the variables. In this section we will construct a CGb with similar properties to the CrGb but  substituting the role of the canonical forms by the coset leaders. From now on, for a given group code $L \trianglelefteq G$,  we will denote by  $\mathrm{CL}(L)$ the  \textit{set of coset leaders } w.r.t. the norm $\left\| . \right\|_{G}$.

\begin{lem}\label{LemaCL} Let  $L \trianglelefteq G$ a group code in $G$, and 
  $\mathbf{l}\in\mathrm{CL}(L)$ and $\mathbf{g}\in G$. If    $g_i=l_i$  for each $i\in \mathrm{supp}(\mathbf{g})$, then  $\mathbf{g}$ is also an element in $\mathrm{CL}(L)$.   
\end{lem} 

Note that the previous result is just a consequence of the definition of  $\left\| . \right\|_{G}$, since if an element has minimum support according to the $G$-norm, then any element whose   support is contained in it must also be of minimal support in the corresponding coset. In the case of binary codes, the same notion of \emph{coset ancestor} is well known, see~\cite[\S11.7]{HP}.

\begin{defin}  Let  $L \trianglelefteq G$ a lattice  code.
A monomial  $x^\mathbf{a}\in [X]$ is an \textit{irredundant term} if for each   $i \in \mathrm{supp}(\mathbf{a})$ such that  $x^\mathbf{a}=x^{\mathbf{b}}x_i$ then  $\mathbf{b} \in \mathrm{CL}(L)$. We will denote by  $\mathrm{IT}(L)$ the set of all irredundant terms.
In other words, 
$$x^\mathbf{a} \in \mathrm{IT}(L) \Leftrightarrow  (x^\mathbf{a}=x^{\mathbf{b}}x_i \Rightarrow \mathbf{b} \in \mathrm{CL}(L) \; \forall x_i \in \mathrm{supp}(x^\mathbf{a})).$$
\end{defin}

\begin{defin}\label{BGCompletaExt}  Let  $L \trianglelefteq G$ a lattice code, and $\prec$ a total ordering. The
\textit{extended complete reduced Gr\"obner  basis} (ECrGb) w.r.t. $\prec$ of a  CGb is the set of binomials in $\mathbb F_2[\mathbf x]$ given by
$$\mathrm{ECrGb}(L)=\{x^\mathbf{a}-x^\mathbf{b} \mid x^\mathbf{a} \in\mathrm{IT}(L), b\in \mathrm{CL}(L), \mathbf{a}\neq \mathbf{b} \}.$$
\end{defin}
 The following procedure computes a $\mathrm{ECrGb}(L)$  w.r.t. a total ordering $\prec$ for the group code $L \trianglelefteq G$ specified by its parity check matrix.
 \begin{algorithm}\label{Alg1} Algorithm for computing  $\mathrm{ECrGb}(L)$\\
 	\textbf{Input} $\prec$ a total ordering,  $n, H$ for a given lattice code $L$.\\
 	\textbf{Output} $\mathrm{G}(I, \prec)$ ECrGb of $L$.
\begin{algorithmic}
	\State $G \gets  \emptyset$,  $List\gets [1]$, $r \gets 0$
	\While{$List \neq \emptyset$} 
	\State $w \gets NextTerm[List]$
	\If{if $w \in IT(L)$}
	\State $c \gets False$,
	\State $v' \gets \xi(w)$,
	\State $( \Lambda , j) \gets Member[v_{0} , {v_{1} , . . . , v_{r}}]$,
	 \If{$\Lambda = True$} 
	\If{$(deg(w _{j1}) \neq 0)$ and $(deg(w)=deg(w _{j1}))$}
	\State $List \gets InsertNext[w, List]$,
	\State $N_{j} \gets  N_{j} \cup \{w\}$,
	\State$c \gets True$,
	\EndIf
	\For{$i = 1$ to $Length [N_{j}]$}
	\State $G \gets G \cup \{w - w _{ji}\}$,
	\If{$c =True$}
	\State  $G \gets G \cup \{w _{ji} - w\}$,
	\EndIf
	\EndFor
	\Else
	\State $r\gets r + 1$,
	\State $v_{r} \gets v'$,
	\State $w_{r1} \gets w$, $N_r\gets \{w_{r1} \}$,
	\State $List \gets InsertNext[w_{r1}, List]$,
	\EndIf
	\EndIf
	\EndWhile 
\end{algorithmic}
\textbf{Return} G
 \end{algorithm}
where \textbf{InsertNext[w,List]} inserts the product $x\mathbf{w}$ for each $x \in \{x_1,\ldots, x_n\}$ in  \textit{List}, a list always ordered in increasing ordering w.r.t. $\prec$. The output of the instruction 
\textbf{Member[$v_{0} , {v_{1} ,\ldots , v_{r}}$]} has as first component TRUE if $v_{0}$ is in the syndrome list $[v_{1} , . . . , v_{r}]$ or FALSE, in the other case. Its second component is the position of the syndrome when it is in the list.  \textbf{N} is the list where we keep the coset leaders meanwhile they are found.
 \textbf{List} is a list where the multiples of the coset leaders are kept in increasing order w.r.t. $\prec$, i.e. the irredundant terms.

  The following result will be needed before it can be proven    the correctness of Algorithm~\ref{Alg1}.

\begin{lem}\label{TeoAlg1}  Let  $L \trianglelefteq G$ a lattice code. If $x^\mathbf{a} \in [X]$ is a monomial, then $\mathbf{a} \in \mathrm{CL}(L) $ if and only if $x^\mathbf{a} \in List$ and  $x^\mathbf{a} \in N$  computed in Algorithm~\ref{Alg1}.
	\end{lem}
	
	\begin{proof} We will use induction over the degree of the monomials in   $[X]$. Let 
		    $x^\mathbf a \in [X]$ be a monomial such that  $\mathbf a \in \mathrm{CL}(L)$ and degree $k+1$ and we will assume that for all the elements $\mathbf b\in \mathrm{CL}(L)$ such that the monomial $x^\mathbf b$ is of degree less or equal to $k$ are also in the sets $List$ and $N$. Note that $x^\mathbf a$ can be expressed as  $x^\mathbf a = x_i x^\mathbf b$, where $x_i \in X$ and $\mathrm{deg}(x^\mathbf b)=k$. By  Lemma~\ref{LemaCL}
		    $\mathbf b \in \mathrm{CL}(L)$ and therefore $x^\mathbf b \in List$. Therefore, the algorithm will insert in
	  $List$ all the degree $k+1$ multiples of  $x^\mathbf b$, in particular $x^\mathbf a = x_i x^\mathbf b$, and hence $x^\mathbf a \in List$. Now, since 
		     $x^\mathbf a \in List$, the algorithm will take it with the procedure $NextTerm$, and as  $\mathbf a \in \mathrm{CL}(L)$, by Lema~\ref{LemaCL}, $x^\mathbf a\in TI(L)$, henceforth  $x^\mathbf a$  will be added to the list  $N$.
		     
		   Now we have to prove that a monomial $x^\mathbf a$ of degree $k+1$ that has been added to  $List$ and $N$ fulfills   $\mathbf a \in \mathrm{CL}(L)$. Note that if  $x^\mathbf a \in List$  also its canonical form $\mathrm{Can}(x^\mathbf a)$ is in  $List$. Note that $\mathrm{Can}(x^\mathbf a)$ is the least element w.r.t. $\prec$  that is in the same coset as $x^\mathbf a$, thus  $\mathrm{Can}(x^\mathbf a) \in N$. Also $\mathrm{Can}(x^\mathbf a)$ corresponds to a coset leader and, as $x^\mathbf a \in N$ and  $\mathrm{Can}(x^\mathbf a) \in N$, thus
		   
		   \begin{enumerate}
		   	\item either $x^\mathbf a=\mathrm{Can}(x^\mathbf a)$, therefore   $x^\mathbf a$  entered in the step ($w_{r1} \gets w$, $N_r\gets \{w_{r1} \}$,) and hence $\mathbf a \in \mathrm{CL}(L)$, 
		   	\item or  $x^\mathbf a\neq \mathrm{Can}(x^\mathbf a)$ , in this case the condition ($(deg(w _{j1}) \neq 0)$ and $(deg(w)=deg(w _{j1}))$) is fulfilled and therefore   $x^\mathbf a$ is added to $N$, hence  $\mathbf a \in \mathrm{CL}(L)$.
		   \end{enumerate}
		   Thus  $\mathbf a \in \mathrm{CL}(L)$ and we conclude the proof.
	\end{proof}

	\begin{thm}  Let  $L \trianglelefteq G$ a lattice code   and a total ordering $\prec$. Algorithm~\ref{Alg1}
	 computes  the   ECrGb w.r.t. $\prec$ for  $L$.
\end{thm}
\begin{proof}
Assume, as induction hypothesis, that the set $G$ in the algorithm and the ECrGb w.r.t. $\prec$ for  $L$ coincide when restricted to those bimomials whose leading terms have degree at most $k$. 

	 If there is no leading term in the ECrGb of degree $k+1$, then all the elements in  $\mathrm{TI}(L)$ of degree $k+1$ correspond to coset with only one coset leader. Moreover, in this case,   $\Lambda=False$ for all the terms of degree  $k+1$.  On the other hand, if there are not leading terms in $G$ of degree  $k+1$  and all the irredundant terms of degree $k+1$  have been included in $List$ (see Lemma~\ref{TeoAlg1}), then we have the same conclusion for the terms in $\mathrm{TI}(L)$ of degree $k+1$ (note that all the elements in $\mathrm{TI}(L)$ of degree $k + 1$ correspond to coset with only one coset leader). Therefore, there will not be any element of degree  $k+1$ in the ECrGb and, in this case,   $G$ and ECrGb coincide in all the binomials whose leading terms have degree at most $k+1$.\\
	Suppose now that there are binomials whose leading term is of degree  $k+1$. Let  $ x^\mathbf a- x^\mathbf b \in$ECrGb and $\mathrm{deg}(x^\mathbf a)=k+1$. Then, by the definition of the ECrGb one has that 
	 $$x^\mathbf a \in \mathrm{TI}(L), \mathbf b\in \mathrm{CL}(\mathbf a),\,\mathbf a\neq \mathbf b.$$   
	 If  $x^\mathbf a \in \mathrm{TI}(L)$, then $x^\mathbf a \in List$ by the definition of irredundant term and one of the two InsertNext steps in Algorithm~\ref{Alg1}  and  $x^\mathbf a$ satisfies step ($w \in \mathrm{IT}(L)$). Also by Lemma~\ref{TeoAlg1}, $\mathbf b\in \mathrm{CL}(\mathbf a)$ implies $x^\mathbf b\in List$ and $x^\mathbf b\in N$.  Therefore the binomial  $x^\mathbf a- x^\mathbf b \in G$ by one of the statemens ($G \gets G \cup \{w - w _{ji}\}$),
	 ($G \gets G \cup \{w _{ji} - w\}$) in the algorithm. Therefore those   binomials in  ECrGb whose leading terms have degree at most $k+1$ are also included in $G$.\\
	 We will check the other inclusion now. Let $ x^\mathbf a- x^\mathbf b \in G$, whose leading term fulfills $deg(x^\mathbf a)=k+1$. Then, $x^\mathbf a\in \mathrm{IT}(L)$, $x^\mathbf b\in {IT}(L)$, and
	 \begin{enumerate}
	 	\item if the inclusion was made at the step ($G \gets G \cup \{w - w _{ji}\}$) then  $x^\mathbf b\in N$ and $\mathbf b\in \mathrm{CL}(\mathbf a)$ (by Lemma~\ref{TeoAlg1}), thus $x^\mathbf a-x^\mathbf b\in$ ECrGb,
	 	\item if the inclusion was made at the step  ($G \gets G \cup \{w _{ji} - w\}$) then  $x^\mathbf a\in N$ and $x^\mathbf b\in N$, and $\mathbf b\in \mathrm{CL}(\mathbf a)$, thus, $x^\mathbf a-x^\mathbf b\in$  ECrGb.
	 \end{enumerate} 
 Hence those   binomials in $G$ whose leading terms have degree at most $k+1$ are also included in   ECrGb and, by induction hypothesis, we have the result. 
\end{proof}

Note that the algorithm termination is clear due the fact that the number of cosets is finite and therefore the bound on the number of elements in $List$.

\section{Decoding using an ECrGb}\label{sec:final}
 
In Theorem~\ref{teop} we find an element $\mathbf c$ such that $\left\| \mathbf a-\mathbf c \right\|_{G}$ is minimal reducing the monomial $x^\mathbf a$ with a Gr\"obner basis  $\mathcal G_{\prec}({L})$. Thus reducing  $x^{-\mathbf{a}}$ w.r.t. $\mathcal G_{\prec}({L})$, we have that 
if  
\begin{equation}\label{eq:eprima}
	\mathbf{e}^{\prime}= \varphi(\mathrm{Can}(x^{-\mathbf{a}},G_{L})),\end{equation}	
	 then $\mathbf{c}^{\prime}=-\mathbf{a}-\mathbf{e}^{\prime}$ is d in $L$ and $\mathbf{c}_{0}=-\mathbf{c}^{\prime}$ is the nearest codeword in $L$ to $\mathbf a$, i.e. $ \left\|\mathbf{c}_{0}-\mathbf{a} \right\|_{G}$ is minimal. In this section we propose an algorithm similar to the one in~ \cite{lat1} but using a ECrGb. 
	 
	 We will slightly abuse the notation and also denote by $\varphi(\mathrm{Can}(x^{\mathbf{a}},\mathrm{ECrGb}(L))$ and  $\varphi(\mathrm{Can}(x^{-\mathbf{a}},\mathrm{ECrGb}(L))$ to the list of reductions of a given momomials by the $\mathrm{ECrGb}(L).$  Also $\mathbf e$ and $\mathbf e^\prime$ will be lists of coset leaders and  $\left\| \mathbf e \right\|_G$, $\left\| \mathbf e^{\prime} \right\|_G$ will be the $G$-norm of any of the elements in the list (note that the $G$-norm of any element in the list is the same since they are coset leaders of the same coset).
	 
	 Then  we prove that the use of this extended set of binomials will provides different optimal solutions w.r.t. the   $l_1$-norm and these solutions are associate to elements   $\mathbf v\in L$ such that $\left\| \mathbf v- \mathbf u \right\|_{l_1}$ is minimal for a given vector $\mathbf u$, where   $\mathrm{d}_{l_{1}}$ is defined as usual, if $\mathbf g_1,\mathbf g_2\in G$,   $\mathrm{d}_{l_{1}}(g_1,g_2)=\sum^{n}_{i=1}| g_{1,i}-g_{2,i} |$.

\begin{algorithm}\label{alg2} Nearest points in a Lattice w.r.t. $l_1$ metric\\
	\textbf{Input} $\mathbf u=(u_1,\ldots,u_n )$, $C(\Delta)$, $P(\Delta), \mathrm{ECrGb}(L)$ \\
	\textbf{Output} $\mathbf v=[(v_1[j],\ldots,v_n[j])]$ a list of nearest points in the  lattice.
	\begin{algorithmic}
		\State $\mathbf e \gets  \emptyset$,  $\mathbf e^\prime \gets  \emptyset$, $\mathbf v \gets  \emptyset$
	\For{$i = 1,\ldots,\mathrm{rank}(\Delta)$ } 
	\State $k_i\gets \left\lfloor u_{i}/\Delta_{W_{i}} \right \rfloor $, $R_i\gets \mathrm{Remainder}(u_i,|\Delta_{W_{i}} |)$, $r_i\gets R_i/|P_{W_{i|}}$,  $z_i\gets \left\lceil r_i \right\rfloor$,
	\EndFor
	\State $\mathbf e\gets  \varphi (Red(x^\mathbf z,\mathrm{ECrGb}(L)))$,
	\If{$\left\| \mathbf e \right\|_G=0$}  $\mathbf c\gets \mathbf z$,
	\ElsIf{ $\left\| \mathbf e \right\|_G=1$} $\mathbf c\gets \mathbf z- \mathbf e$
	\Else 
	\State ${\mathbf e^{\prime} \gets \varphi(Red(x^{- \mathbf z}, \mathrm{ECrGb}(L))), \mathbf c_0=\mathbf  z+ \mathbf e^{\prime}}$,
	\If{$\left\| \mathbf e \right\|_G < \left\| \mathbf e^{\prime} \right\|_G$}  $\mathbf c\gets \mathbf  z- \mathbf e$
	\Else
	\If{$\left\| \mathbf e \right\|_G = \left\| \mathbf e^{\prime} \right\|_G$}
	\If{ $d_{l_1} (\mathbf z,\mathbf c) \leq d_{l_1} (\mathbf  z,\mathbf  c_0 )$} $\mathbf c\gets \mathbf z- \mathbf e$
	\Else  $\quad\mathbf  c\gets \mathbf  c_0$
	\EndIf
	\Else \quad $\mathbf c\gets \mathbf c_0$
	\EndIf
	\EndIf
	\EndIf
		\For{$j = 1,\ldots,\mathrm{length}(\mathbf v)$} 
	\For{$i = 1,\ldots,rank(\Delta)$} 
	\State $v_i[j] \gets k_i |\Delta_{W_{i}}| + c_i[j] |P_{W_{i}}|$
	\EndFor
	\EndFor
	\end{algorithmic}
	\textbf{Return} $\mathbf v$
\end{algorithm}

 Let fix the ECrGb  for a given ordering. Note that  if $\mathbf g\in G$ is an arbitrary vector and one  computes $\mathbf e=\varphi(\mathrm{Can}( x^\mathbf g,ECrGb ))$ there could be several candidates since there are several canonical forms as coset leaders, hence there could be several candidates $\mathbf e$ as we will show in the examples, and henceforth several values for $\mathbf c=\mathbf g -\mathbf e$ that are nearest codewords w.r.t. $\left\|.\right\|_G$. The same situation could happen with the values for $\mathbf e^\prime$ in Equation~(\ref{eq:eprima}) and hence, we could get several candidates $\mathbf c_0$. Thus, we can get more than one decoded vector with Algoritm~\ref{alg2} meanwhile the procedure in Theorem~\ref{teop} provides only one.
 
 Note that in any case, the vectors provided by   Algoritm~\ref{alg2} are optimal solutions w.r.t. the  $l_1$-norm ($\left\| v-u \right\|_{l_1}$), henceforth we can talk about  candidate vectors $\mathbf v$ such that  $\left\| \mathbf v-\mathbf u \right\|$ is minimum. The previous discussion proves the following result.

 \begin{thm}  Let  $L \trianglelefteq G$ a lattice code and $\mathbf u\in \mathbb R^n$. Algorithm~\ref{alg2}
 	computes  the   list $\mathbf v$ of vectors in $L$ that minimize the ${l_1}$ distance with $\mathbf u$, an they are candidates to minimize the ordinary $l_2$ norm.
 \end{thm}

\subsection{Examples}\label{ejem} This examples were run in a computer with 
  Gateway CPU: 160GHz and RAM: 3.00 GB,using the computer algebra package GAP~\cite{GAP}  version 4.7.6. 

\begin{example}\label{ex:1} Consider the lattice 
	 $A_4$ given by the matrix (see \cite[\S V.C]{lat3})
	  $$B=\left(\begin{array}{rrrrr}
	 	-1&1&0&0&0\\
	 	0&-1&1&0&0\\
	 	0&0&-1&1&0\\
	 	0&0&0&-1&1
	 \end{array}\right).
	 $$  
	  We have that the group is $G=\mathbb{Z}_2\times \mathbb{Z}_6\times\mathbb{Z}_{12}\times\mathbb{Z}_4$ and the cross-projections of the lattice are given by
	 $$P(\Delta)=\mathrm{diag}( 1/2\sqrt{2} ,  1/6\sqrt{6} , 1/12\sqrt{12} ,1/2\sqrt{5}) \hbox{ and }C(\Delta)=\mathrm{diag}(\sqrt{2} ,\sqrt{6}, \sqrt{12} ,2\sqrt{5}).$$
	   We have used the degree-lexicographical ordering $\prec$ with four variables. If we compute the ECrGb  we have that the monomials 
		$x_1 x_2 x_3^2, \,  x_3 x_4^3$   and   $x_2^2 x_4^2$
	have the same syndrome (i.e. they lay in the same coset) and have same total degree which is minimal in their coset. Thus following   Definition~\ref{BGCompletaExt} the binomials  
		$x_1 x_2 x_3^2-x_3 x_4^3,  x_3 x_4^3-x_1 x_2 x_3^2,   x_1 x_2 x_3^2-x_2^2 x_4^2,  x_2^2 x_4^2-x_1 x_2 x_3^2,  x_3 x_4^3-x_2^2 x_4^2,  x_2^2 x_4^2-x_3 x_4^3$
	are in the ECrGb.
	
	Let $\mathbf u=(5.12,6.3,54,63)$, we have used   Algorithm~\ref{alg2} with the  ECrGb to compute the nearest lattice points to $\mathbf u$. Then we have that  $\mathbf z=(1,3,7,0)$, and therefore  $x^\mathbf z=x_1 x_2^3 x_3^7$. If we compute its canonical form $$\mathrm{Can}( x^\mathbf z,ECrGb )=\{ x_3 x_4^2, \, x_2^2 x_4 \},$$ thus $\mathbf e=\{ (0,0,1,2);(0,2,0,1)\}$, and  $ \left\|{\mathbf e}[i]  \right\|_G=3$ for $i=1,2$. Now if we compute  $x^{-\mathbf z}=x_1 x_2^3 x_3^5$ we have that  $$\mathrm{Can}(x^{-\mathbf z}, GCe )=\{ x_2^2 x_3, \, x_3^2 x_4 \},$$ and 
 $\mathbf e^{\prime}=\{ (0,2,1,0);(0,0,2,1)\}$, $\mathbf c_0=\{ (1,5,8,0);(1,3,9,1) \}$. Now we will compute the values for $\mathbf c$.
 \begin{itemize}
 	\item As  $ \left\|{\mathbf e}[1]   \right\|_G= \left\|{\mathbf e^\prime}[1]   \right\|_G=3$ we compare the  $l_1$ norms 
 	$$\mathbf c=\mathbf z-{\mathbf e}[1] =(1,3,6,2) \hbox{ and } \mathrm{d}_{l_1} (\mathbf z,\mathbf c)=\mathrm{d}_{l_1} (\mathbf z,\mathbf c_0 )=3.$$ Thus  $\mathbf c_1=\mathbf c=(1,3,6,2)$.
 	\item In the second case,  $ \left\|{\mathbf e}[1]   \right\|_G= \left\|{\mathbf e^\prime}[2]   \right\|_G=3$ again we compare the norms
 		$$\mathbf c=\mathbf z-{\mathbf e}[1] =(1,3,6,2) \hbox{ and } \mathrm{d}_{l_1} (\mathbf z,\mathbf c)=\mathrm{d}_{l_1} (\mathbf z,\mathbf c_0 )=3.$$ And $\mathbf c_2=\mathbf c=(1,3,6,2)$.
 		\item Now, for  $ \left\|{\mathbf e}[2]   \right\|_G= \left\|{\mathbf e^\prime}[2]  \right\|_G=3$ and $\mathrm{d}_{l_1} (\mathbf z,\mathbf c)=5>\mathrm{d}_{l_1} (\mathbf z,\mathbf c_0 )=3$, thus  $\mathbf c_3= \mathbf c_{0_{1}}=(1,5,8,0)$.
 		\item In the last case, $ \left\|{\mathbf e}[2]  \right\|_G= \left\|{\mathbf e^\prime}[2]   \right\|_G=3$ and again we have  $\mathrm{d}_{l_1} (\mathbf z,\mathbf c)=5>\mathrm{d}_{l_1} (\mathbf z,\mathbf c_0 )=3$, hence $\mathbf c_4=\mathbf c_{0_{2}}=(1,3,9,1)$.
\end{itemize} 
We got three vectors nearest to $\mathbf z$, that correspond to nearest lattice points to $\mathbf{u}$ w.r.t. the $G$-norm.
	\begin{center}
		$\{ ( 4.95 ,6.12 ,53.69 ,64.85 ),( 4.95 ,6.94 ,54.27 ,62.61 ),( 4.95 ,6.12 ,54.56 ,63.73 )\}$.
	\end{center}
The nearest one, the one achieving the least value for the  $\left\|\cdot \right\|$-norm,  is the second one.
\end{example}

\begin{example} Consider the check-board lattice 
  $D_4$ with matrix $B$ given by 
  $$B=\left(\begin{array}{cccc}
  	1&1&0&0\\
  		1&0&1&0\\
  			1&1&1&1\\
	2&0&0&0
  \end{array}\right).
$$  
  It can be checked that in this case $G=\mathbb Z_2\times\mathbb Z_6\times \mathbb Z_6\times \mathbb Z_2$ (see \cite[Example 4]{lat1}) and the cross-sections are 
  
  $$P(\Delta)=\mathrm{diag}( 1/\sqrt{2},  1/\sqrt{6},  1/\sqrt{3},  1)\hbox{ and } C(\Delta)=\mathrm{diag}(\sqrt{2},   \sqrt{6},   2\sqrt{3},  2).$$ 
  Again, 
   we have computed the ECrGb w.r.t. the  degree-lexicographic ordering $ \prec$.
   Consider we get  $\mathbf u=(56.12,46.3,54,63)$. We have that $\mathbf z=(1,5,4,0)$ and $x^\mathbf z=x_1 x_2^5 x_3^4 x_4$. If we compute its canonical form 
   $$\mathrm{Can}( x^\mathbf z,\mathrm{ECrGb})=\{ x_1 x_2 x_4,x_3^2 x_4, x_2^2 x_3  \},$$
   
   we have that  $\mathbf e=\{ (1,1,0,1);(0,0,2,1)  ,(0,2,1,0)  \}$, and  $\left\| \mathbf e_i \right\|_G=3$ for $i=1,2,3$. Now, $x^{-\mathbf z}=x_1 x_2 x_3^2 x_4$ and  $\mathrm{Can}(x^{-\mathbf z},\mathrm{ECrGb})=\{ x_3  \}$, therefore  $\mathbf e^{\prime}={ (0,0,1,0)  }$ and  $\mathbf c_0={ (1,5,5,1)  }$. Since   $\left\| {\mathbf e}[i]  \right\|_G> \left\|\mathbf e^{\prime} \right\|_G$ for each $i=1,2,3$, then $\mathbf c=\mathbf c_0=(1,5,5,1)$. In this case, we get only one candidate as the nearest lattice point to $\mathbf u$ given by  $ ( 55.86 ,46.13 ,54.85 ,63 ) $.
\end{example}
\section{Conclusions}
In this work  we have introduced an algorithm that computes the extended complete reduced Gr\"obner  basis of a lattice code w.r.t. a given total degree compatible ordering. We also show a decoding algorithm that uses the reduction process provides by that extended basis which guaranties that one gets all the closest vectors to the original one w.r.t. the $G$-norm and, therefore, providing all the possible candidates w.r.t. the $G$-norm for the solution of the Close Vector Problem associated to the lattice.

\section*{Aknowledgements} First and second authors are partially supported by International Funds and Projects Management Office (Cuba) under the code
PN223LH010-024. Fourth author is  supported by Grant TED2021-130358B-I00 funded by MCIN/AEI/10.13039/ 501100011033 and by the “European Union NextGenerationEU/PRTR”.

\bibliographystyle{plain} 
\bibliography{BGComp}  
\end{document}